\documentstyle[aps,preprint,tighten,12pt]{revtex}
\begin{document}
\draft
\title{Contact interaction in the nonrelativistic pion absorption operator}
\author{D.A.Zaikin and I.I.Osipchuk}
\address{Institute for Nuclear Research of Russian Academy of Sciences,\\ 117312
 Moscow, Russia}

\maketitle

 \begin{abstract}
Chirally invariant Lagrangian containing nucleon, pion and chiral scalar fields
is considered. A consistent scheme is proposed for obtaining the nonrelativistic
operator of pion absorption by a $NN$-pair taking into account the chiral
scalar exchange. An important role of the contact interaction is pointed out.

\bigskip


\end{abstract}


1. Taking into consideration a scalar meson in the effective chiral
Lagrangian has led to a successful description of the three-nucleon interaction
and pion production in the $NN$-collisions near threshold \cite{1,2}. It was
shown \cite{1} that cross sections of the reactions $pp\to pp\pi^0$, $pp\to
d\pi^+$ and $pp\to pn\pi^+$ near threshold can be explained taking into account
a contact interaction described by the $\pi SNN$-vertex ($S$ being a chiral
scalar-isoscalar meson using as a model approximation for the two-pion exchange
between nucleons).

The Lagrangian describing interaction of the nucleon, pion and chiral scalar
fields can be presented \cite{2} as follows
\begin{equation}
{\cal L}^{ps}_N=\overline{N}i\gamma_{\mu}\partial^{\mu}-g\overline{N}(\sigma+
i\gamma_5\bbox{\tau}\bbox{\varphi})N-\frac{g_s}{f_{\pi}}
S\overline{N}(\sigma+i\gamma_5\bbox{\tau}\bbox{\varphi})N+
{\cal L}_0+{\cal L}_{sb},
\label{1}
\end{equation}
$$
{\cal L}_{sb}=f_{\pi}\mu^2\sigma=f_{\pi}\mu^2\sqrt{f_{\pi}^2-\bbox{
\varphi}^2},
$$
$$
{\cal L}_0=\frac{1}{2}(\partial_{\mu}S\partial^{\mu}S-m_s^2S^2)+U(S)+\frac{1}{2}
(\partial_{\mu}\bbox{\varphi}\partial^{\mu}\bbox{\varphi}+
\partial_{\mu}\sigma\partial^{\mu}\sigma),
$$
$N$, $\varphi$ and $S$ being nucleon, pion and chiral scalar fields,
${\cal L}_{sb}$ being a part of the Lagrangian breaking the chiral symmetry,
$f_{\pi}$ --- the pion decay constant, $g$ and $g_s$ --- $\pi NN$ and $SNN$
coupling constants; $m=gf_{\pi}$ is a nucleon mass, while $\mu$ and $m_s$
are pion and chiral scalar masses respectively; $U(S)$ describes the
self-interactions of the scalar field. Lagrangian (1) describing
the $\pi NN$-interaction is chirally invariant except for ${\cal L}_{sb}$:
at the same time it is nonlinear since pion fields are transformed in a
nonlinear way by chiral transformations while nucleon fields of this
Lagrangian are transformed linearly.

Let us apply a unitary transformation $u(i\gamma_5\bbox{\tau}\bbox{\varphi})$
to the nucleon fields of Lagrangian (1):
$$
N=u(i\gamma_5\bbox{\tau}\bbox{\varphi})\psi\,,\quad u(\sigma
+i\gamma_5\bbox{\tau}\bbox{\varphi})u=f_{\pi}\,.
$$
It is easy to show that
\begin{equation}
u=\sqrt{\frac{f_{\pi}+\sigma}{2f_{\pi}}}\left(1-
i\gamma_5\frac{\bbox{\tau}\bbox{\varphi}}{f_{\pi}+\sigma}\right)\,.
\label{2}
\end{equation}
As a result instead of pseudoscalar $\pi NN$-coupling we obtain pseudovector
coupling and Lagrangian (1) can be presented as follows
\begin{equation}
{\cal L}^{pv}_{\psi}=\overline{\psi}i\gamma_{\mu}{\cal D}^{\mu}\psi-
m\overline{\psi}\psi+
\frac{g}{2m}\overline{\psi}\gamma_{\mu}\gamma_5\bbox{\tau}{\cal D}_{\mu}
\bbox{\varphi }\psi-g_sS\overline{\psi}\psi+{\cal L}'_0+{\cal L}_{sb}\,,
\label{3}
\end{equation}
$$
{\cal
L}'_0=\frac{1}{2}(\partial_{\mu}S\partial^{\mu}S-m_s^2S^2)+U(S)+\frac{1}{2}({\cal
D}^{\mu}\bbox{\varphi})^2\,,
$$
${\cal D}^{\mu}\bbox{\varphi}$ and ${\cal D}^{\mu}\psi$ being covariant
derivatives of the pion and nucleon fields:
$$
{\cal D}^{\mu}\bbox{\varphi}=\partial^{\mu}\bbox{\varphi}-\frac{1}{\sigma+
f_{\pi}}\partial^{\mu}\sigma\bbox{\varphi}\,,
\quad {\cal D}_{\mu}\psi=\left[\partial_{\mu}+i\frac{1}{f_{\pi}(\sigma+
f_{\pi})}\frac{\bbox{\tau}}{2}(\bbox{\varphi}\times\partial^{\mu}\bbox{\varphi})
\right]\psi\,.
$$
Since the transformation connecting $\psi$ and $N$ is a canonical one,
corresponding $T$-matrix elements calculated using Lagrangians (1) and (3)
will be the same for particles on the mass shell. Of couse, both Lagrangians
satisfy PCAC condition
$$
\partial^{\mu}A_{\mu}^{\alpha}=f_{\pi}\mu^2\varphi^{\alpha}\,,
$$
$A_{\mu}^{\alpha}$ being a component of the axial current calculated by use
of (1) or (3).

Expanding (1) and (3) up to terms of the order of  $(\varphi/f_{\pi})^2$ one
obtains
$$
{\cal L}^{ps}_N=\overline{N}(i\not\partial-m)N+\frac{1}{2}\left((\partial_{\mu}
\bbox{\varphi})^2-\mu^2\bbox{\varphi}^2\right)-g\overline{N}i\gamma_5\bbox{\tau}
\bbox{\varphi}N+\frac{g}{2f_{\pi}}\overline{N}N\bbox{\varphi}^2-
$$
\begin{equation}
\frac{g_s}
{f_{\pi}}S\overline{N}\left(f_{\pi}-\frac{\bbox{\varphi}^2}{2f_{\pi}}+i\gamma_5
\bbox{\tau}\bbox{\varphi}\right)N+\frac{1}{2}(\partial_{\mu}S\partial^{\mu}S-
m_s^2S^2)+U(S),
\label{4}
\end{equation}
$$
{\cal L}^{pv}_{\psi}=\overline{\psi}(i\not\partial-
m)\psi+\frac{1}{2}\left((\partial_{\mu}\bbox{\varphi})^2-\mu^2\bbox{\varphi}^2\right)+
\frac{g}{2m}\overline{\psi}\gamma_{\mu}\gamma_5\bbox{\tau}\psi\partial^{\mu}
\bbox{\varphi}-
$$
\begin{equation}
\frac{1}{(2f_{\pi})^2}\overline{\psi}\gamma_{\mu}\bbox{\tau}\psi
(\bbox{\varphi}\times\partial^{\mu}\bbox{\varphi})-g_sS\overline{\psi}\psi+
U(S)+\frac{1}{2}(\partial_{\mu}S
\partial^{\mu}S-m_s^2S^2)\,.
\label{5}
\end{equation}

To describe  the $\pi N\to \pi N)$ amplitude one has to use only parts of
Lagrangians (4) and (5) responsible for the interaction in $\pi N$ sector,
namely
\begin{equation}
{\cal L}^{ps}_{\pi N}=-g\overline{N}i\gamma_5\bbox{\tau}\bbox{\varphi}N+
\frac{g}{2f_{\pi}}\overline{N}N\bbox{\varphi}^2\,,
\label{6}
\end{equation}
\begin{equation}
{\cal L}^{pv}_{\pi N}=\frac{g}{2m}\overline{\psi}\gamma_{\mu}\gamma_5
\bbox{\tau}\psi\partial^{\mu}\bbox{\varphi}-\frac{1}{(2f_{\pi})^2}
\overline{\psi}\gamma_{\mu}\bbox{\tau}
\psi(\bbox{\varphi}\times\partial^{\mu}\bbox{\varphi })\,.
\label{7}
\end{equation}

The $\pi N$ scattering amplitude calculated to the second order of the
perturbation theory (using
Lagrangians (6) and (7)) may be presented by diagrams of Fig.1. Calculations
using (6) and (7)
give the same result as it was expected (see e.g. \cite{3}).

Another consequence of the Lagrangians (4) and (5) equivalency is the amplitude
$\pi N\to SN$
equality being calculated for diagrams of Fig.2 using the interaction
Lagrangians
\begin{equation}
{\cal
L}^{pv}=\frac{g}{2m}\overline{\psi}\gamma_{\mu}\gamma_5\bbox{\tau}
\psi\partial^{\mu}
\bbox{\varphi}-g_sS\overline{\psi}\psi\,,
\label{8}
\end{equation}
\begin{equation}
{\cal L}^{ps}=-
\frac{g_s}{f_{\pi}}S\overline{N}(f_{\pi}+i\gamma_5\bbox{\tau}\bbox{\varphi})N-
g\overline{N}i\gamma_5\bbox{\tau}\bbox{\varphi}N\,.
\label{9}
\end{equation}
The $\pi N\to SN$ amplitude calculated using Lagrangian (8) (pseudovector
coupling) and
diagrams of Fig.2 is as follows
\begin{equation}
T=i\frac{g}{2m}g_s \overline{u}(p')\tau_{\alpha}\left(\frac{\not p_d+m}
{p_d^2-m^2}\not q\gamma_5+
\not q\gamma_5\frac{\not p_x+m}{p_x^2-m^2}\right)u(p)\,,
\label{10}
\end{equation}
$p_d=p+q$\,, $p_x=p'-q$\,, $\tau_{\alpha}$ is a component of the isospin
operator $\bbox{\tau}$.

Let us transform the amplitude (10) by means of the identities
$$
S_F(p_d)\not q\gamma_5u(p)=[1+2mS_F(p_d)]\gamma_5u(p),\quad u(p')\not q
\gamma_5S_F(p_x)=u(p')\gamma_5[1+2mS_F(p_x)],
$$
where $S_F(p)=(\not p+m)/(p^2-m^2)$ is a nucleon propagator, $u(p)$ is a Dirac
free spinor. Such a transformation leads to the following expression for
amplitude $T$:
\begin{equation}
T=ig_sg\overline{u}(p')\tau_{\alpha}\left(S_F(p_d)\gamma_5+\gamma_5S_F(p_x)+
\frac{\gamma_5}{m}\right)u(p),
\label{11}
\end{equation}
which is identical to the expression one obtains in the calculation of the
amplitudes of Fig. 2 using Lagrangian (9) (see also \cite{2}).

\bigskip

2. Normally the pion creation operator for processes $NN\to NN\pi$is being
calculated starting from an effective Lagrangian describing interaction
of pion, nucleon and other fields. As a rule
in such an approach influence of $NN$ interaction on the precise form of
creation operator is not being taken into consideration.
At the same time the pion creation operator depends on the
nucleon potential \cite{4,5,6} when one uses a nonrelativistic equation
to describe nuclear dynamics. Partial conservation of axial current
(PCAC) also connects concrete form of the pion creation (absorption) operator
with nonrelativistic dynamics of nucleon interaction \cite{7}. The
precise form of such an operator for a nonrelativistic case may be obtained
starting from relativistic $T$-matrix, and the final result may turn out
to be  Galilean-invariant \cite{8}.

Let us consider the pion absorption amplitude on a nucleon caused
by the scalar-isoscalar interaction described by Lagrangian as followed
$$
{\cal L}_{int}=-ig\overline N\gamma_5\bbox{\tau}\bbox{\varphi}N-
V_s\overline{N}N.
$$
This amplitude will correspond to diagrams $a$ and $b$ of Fig.2 with
a virtual $S$-meson and will look as two first terms of (11):
\begin{equation}
T_s=igv_s(\bbox{k})\overline{u}(p')\tau_{\alpha}\left(S_F(p_d)\gamma_5+
\gamma_5S_F(p_x)\right)u(p),
\label{12}
\end{equation}
$v_s(\bbox{k})$ being a vertex function, i.e. Fourier transform of the
potential $V_s$. To find the nonrelativistic limit of eq. (12) let us divide
the nucleon propagator $S_F$ into positive and
negative frequency parts\cite{1}:
$$
S_F(p)=\frac{m}{E_p}\left[\frac{\Lambda^+(\bbox{p})}{p_0-E_p}+
\frac{\Lambda^-(-\bbox{p})}{p_0+E_p}\right]=S_F^+(p)+S_F^-(p),
$$
$$
S_F^+(p)=\frac{1}{2E_p}\left[\frac{\not p+m}{p_0-E_p}-\gamma_0\right],\quad
S_F^-(p)=\frac{1}{2E_p}\left[\gamma_0-\frac{\not p+m}{p_0+E_p}\right],
\quad E_p^2=p^2+m^2,
$$
$\Lambda_+(\bbox{p})$ and $\Lambda_-(\bbox{p})$ being projection operators
for positive and negative energy solutions of Dirac equation. Taking advantage
of an expression describing
nonrelativistic limit of $\gamma_5$-matrix, namely
$$
\overline{u}(\bbox{p}_d)\gamma_5u(\bbox{p})\approx\frac{1}{2m}
\left[\bbox{\sigma}(\bbox{p}-\bbox{p}_d)+\bbox{\sigma}(\bbox{p}_d+
\bbox{p})\frac{\varepsilon_{p_{d}}-\varepsilon_p}{4m}\right],
$$
$\varepsilon_p$ being kinetic energy of the nucleon, one can obtain
an expression for the amplitude corresponding to diagram $a$ of Fig. 2:
\begin{equation}
T_d\approx i\frac{g}{2m}v_s(\bbox{k})\chi_f^+\tau_{\alpha}
\left[\frac{-\bbox{\sigma}\bbox{q}+\omega\bbox{\sigma}(\bbox{p}_d+
\bbox{p})/4m}{\varepsilon_p+\omega-\varepsilon_{pd}}-\\
\frac{\bbox{\sigma}(\bbox{p}_d+\bbox{p})}{4m}+
\frac{\bbox{\sigma}(\bbox{p}+\bbox{p}'+\bbox{q}}{2m}\right]\chi_i,
\label{13}
\end{equation}
$\chi$ being a two-component spinor, $\omega$ is the pion energy.

The amplitude $T_x$ of the pion absorption by the nucleon corresponding
to diagram $b$ of Fig. 2 to the same approximation is as follows:
\begin{equation}
T_x\approx i\frac{g}{2m}v_s(\bbox{k})\chi_f^+\tau_{\alpha}
\left[\frac{-\bbox{\sigma}
\bbox{q}+\omega\bbox{\sigma}(\bbox{p'}+\bbox{p}_x)/4m}{\varepsilon_{p'}-\omega-
\varepsilon_{px}}+\\
\frac{\bbox{\sigma}(\bbox{p}_x+\bbox{p'})}{4m}-\frac{\bbox{\sigma}(\bbox{p}+
\bbox{p'}-\bbox{q}}{2m}\right]\chi_i.
\label{14}
\end{equation}
Taking into consideration that $\varepsilon_{p'}=\varepsilon_p+\omega$
one can see that the first terms of the expressions (13) and (14) correspond
to a matrix element of the interaction operator
$$
H(x)=-\frac{g}{2m}\bbox{\sigma}\bbox{\nabla}\varphi_{\alpha}-
\frac{g}{8m^2}\{\bbox{\sigma}\bbox{p},\,\dot {\varphi}_{\alpha}\},
$$
$$
\dot {\varphi}_{\alpha}=-i\omega\varphi_{\alpha},\quad
\varphi_{\alpha}=\tau_{\alpha}\exp(i\bbox{q}\bbox{x}),
$$
$\{A,\,B\}$ is an anticomutator . This matrix element is taken between
nonrelativistic nucleon wave functions calculated to the first approximation
concerning the potential $V_s$. The rests of expressions (13) and (14) are
so-called contact terms presented by Fig. 3.

The contribution of contact terms to the absorption amplitude is equal to
\begin{equation}
T_{c}=-i\frac{g}{2m}v_{s}(\bbox{k})\chi_f^+\tau_{\alpha}
\frac{\bbox{\sigma}(\bbox{p}-\bbox{p'}-\bbox{q})}{2m}\chi_i,
\label{15}
\end{equation}
and interaction $H_c(x)$ corresponding to (15) in the coordinate space has
a form:
$$
H_c(x)=\frac{g}{4m^2}\bbox{\sigma}(\varphi_{\alpha}\bbox{\nabla}V_s+
2V_s\bbox{\nabla}\varphi_{\alpha});
$$
$$
T_c=\int\exp (-i\bbox{p'}\bbox{x})H_c(x)\exp (\bbox{p}\bbox{x})\,dx.
$$
Total contribution of $H(x)+H_c(x)$ to the pion absorption operator
coincides with leading terms of the expression for nonrelativistic
pseudoscalar $\pi N$ interaction obtained by means of
Foldy--Wouthuysen transformation for a nucleon moving in a scalar potential
\cite{6}. It is possible to show that a similar procedure leads to the same
result for leading terms in the case of pseudoscalar $\pi N$ interaction for
a nucleon moving in a potential being transformed as the forth component
of a four-vector.

While considering processes of pion absorption (or creation) by a nucleon pair
a static propagator of exchange is often used. Such a propagator takes into
account nucleon correlations and this is quite reasonable when particles heavier
than the pion are involved in the process under consideration. Considering
the $S$-scalar contribution ,e.g. for reaction $pp\to d\pi^+$
near threshold, the authors of \cite{1} took into account $Z$-diagrams
for the $S$-scalar exchange between nucleons, and also the contact term
in the $SN\to\pi N$ amplitude as it follows from Lagrangian (9).
The contribution of the latter turned out to be more than one order
bigger than the contribution of $Z$-diagrams.

Note that in case of the $S$-scalar exchange one should take into account
a contribution of the absorption (or creation) operator arising in
the nonrelativistic description of the internucleonic
interaction. As it is seen from (15) and (11) , near threshold , where
the pion momentum is small, the contribution of leading contact $\pi SNN$ term
must be diminished by factor 1/2 since
$$
\overline{u}(\bbox{p'})\frac{\gamma_5}{m}u(\bbox{p})\approx
\frac{\bbox{\sigma}(\bbox{p}-\bbox{p'})}{2m^2},
$$
while the contribution of $T_c$ is proportional
to $-\bbox{\sigma}(\bbox{p}-\bbox{p'})/4m^2$.

It seems interesting to estimate a contribution of the chiral scalar from the
term
$$
-g_sS\overline{N}(f_{\pi}-\varphi^2/2f_{\pi}+i\bbox{\tau}\bbox{\varphi}
\gamma_5)N/f_{\pi}
$$
of Lagrangian (4) to
the amplitudes of reactions $NN\to NN\pi\pi$ near threshold where as it was
shown e.g. in \cite{9} it is necessary to
take into account excitation of Roper $N_{11}$-resonance.

\newpage

\centerline{FIGURE CAPTIONS}

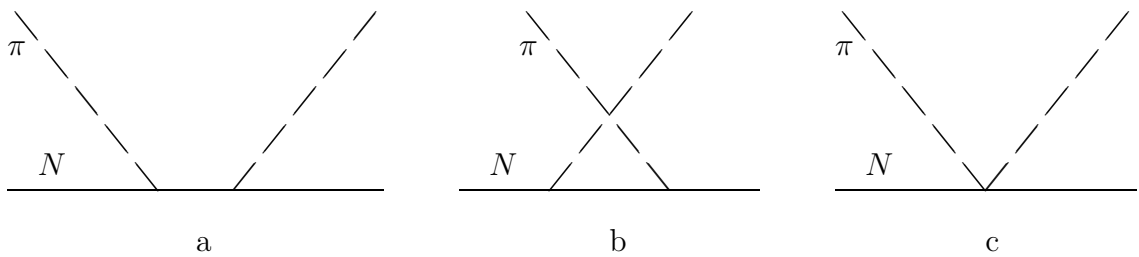
Fig. 1. $\pi N$ scattering amplitudes.

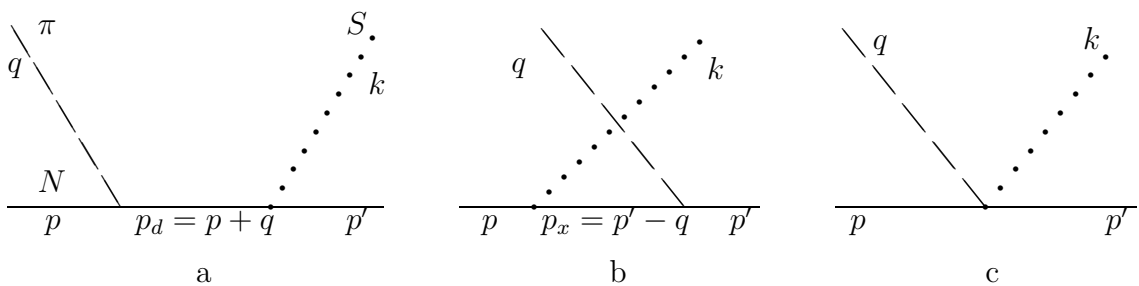
Fig. 2. $\pi N\to SN$ amplitudes: a, b --- for pseudovector interaction, a, b,
c --- for pseudoscalar interaction.

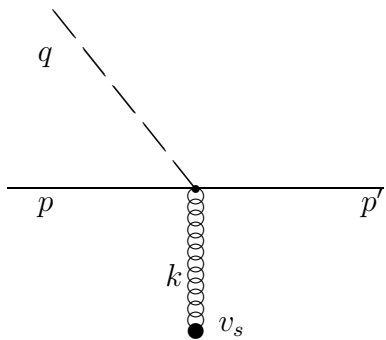
Fig. 3. Contact terms of the pion absorption operator.

\newpage

\begin{picture}(150,50)
\put(0,20){\line(1,0){50}}
\multiput(20,20)(-4,5){5}{\line(-4,5){3}}
\multiput(30,20)(4,5){5}{\line(4,5){3}}
\put(4,22){$N$}
\put(0,38){$\pi$}
\put(25,12){a}

\put(60,20){\line(1,0){40}}
\multiput(72,20)(4,5){5}{\line(4,5){3}}
\multiput(88,20)(-4,5){5}{\line(-4,5){3}}
\put(64,22){$N$}
\put(68,38){$\pi$}
\put(80,12){b}
\put(73,4){Fig. 1}

\put(110,20){\line(1,0){40}}
\multiput(130,20)(-4,5){5}{\line(-4,5){3}}
\multiput(130,20)(4,5){5}{\line(4,5){3}}
\put(114,22){$N$}
\put(110,38){$\pi$}
\put(130,12){c}
\end{picture}

\begin{picture}(150,60)
\put(0,20){\line(1,0){50}}
\multiput(15,20)(-3,5){5}{\line(-3,5){2.5}}
\multiput(35,20)(1.5,2.5){10}{\circle*{0.7}}
\put(4,22){$N$}
\put(5,17){$p$}
\put(4,43){$\pi$}
\put(17,17){$p_d=p+q$}
\put(45,17){$p'$}
\put(45,43){$S$}
\put(48,35){$k$}
\put(0,38){$q$}
\put(25,10){a}

\put(60,20){\line(1,0){40}}
\multiput(90,20)(-4,5){5}{\line(-4,5){3}}
\multiput(70,20)(2,2){12}{\circle*{0.7}}
\put(63,17){$p$}
\put(67,38){$q$}
\put(93,37){$k$}
\put(96,17){$p'$}
\put(71,17){$p_x=p'-q$}
\put(80,10){b}
\put(76,4){Fig. 2}

\put(110,20){\line(1,0){40}}
\multiput(130,20)(-4,5){5}{\line(-4,5){3}}
\multiput(130,20)(2,2.5){9}{\circle*{0.7}}
\put(115,41){$q$}
\put(112,17){$p$}
\put(146,17){$p'$}
\put(143,41){$k$}
\put(130,10){c}
\end{picture}

\begin{picture}(50,70)
\put(0,40){\line(1,0){50}}
\multiput(25,40)(-4,5){5}{\line(-4,5){3}}
\put(25,40){\circle*{1}}
\multiput(25,39)(0,-1.5){12}{\circle{2}}
\put(4,37){$p$}
\put(4,57){$q$}
\put(47,37){$p'$}
\put(21,27){$k$}
\put(28,21){$v_s$}
\put(25,21){\circle*{2}}
\put(21,10){Fig. 3}
\end{picture}

\end{document}